\documentclass[12pt] {IEEEtran}
\onecolumn
\usepackage{setspace}
\newtheorem{prop}{Proposition}

\newtheorem{cor}{Corollary}

\newtheorem{lm}{Lemma}

\newtheorem{thm}{Theorem}

\newcommand{\be}{\begin{eqnarray}}
\newcommand{\ee}{\end{eqnarray}}
\newcommand{\benn}{\begin{eqnarray*}}
\newcommand{\eenn}{\end{eqnarray*}}
\def\IR{\rm I \kern-0.20em R}
\newcommand{\utwi}[1]{\mbox{\boldmath $ #1$}}

\newcommand{\bthm}{\begin{thm}}
\newcommand{\ethm}{\end{thm}}

\newcommand{\bcor}{\begin{cor}}
\newcommand{\ecor}{\end{cor}}
\newcommand{\bprop}{\begin{prop}}
\newcommand{\eprop}{\end{prop}}
\newcommand{\blm}{\begin{lm}}
\newcommand{\elm}{\end{lm}}
\newcommand{\beq}{\begin{equation}}
\newcommand{\eeq}{\end{equation}}
\newcommand{\ber}{\begin{eqnarray}}
\newcommand{\eer}{\end{eqnarray}}

\newcommand{\bproof}{\begin{proof}}
\newcommand{\eproof}{\end{proof}}

\newcommand{\diag}{\mathop{\mbox{\rm diag}}}

\newcommand{\bit}{\begin{itemize}}
\newcommand{\eit}{\end{itemize}}
\newcommand{\ben}{\begin{enumerate}}
\newcommand{\een}{\end{enumerate}}
\newcommand{\bdesc}{\begin{description}}
\newcommand{\edesc}{\end{description}}
\newcommand{\beqarrn}{\begin{eqnarray*}}
\newcommand{\eeqarrn}{\end{eqnarray*}}
\newcommand{\bproofof}{\begin{proofof}}
\newcommand{\eproofof}{\end{proofof}}
\newenvironment{rem}{\begin{trivlist}\item[]{\bf
Remark:}\hspace{4mm}}{\end{trivlist}}
\newcommand{\brem}{\begin{rem}}
\newcommand{\erem}{\end{rem}}
\newenvironment{rems}{\begin{trivlist}\item[]{\bf
Remarks}\begin{itemize}}{\end{itemize}\end{trivlist}}
\newcommand{\brems}{\begin{rems}}
\newcommand{\erems}{\end{rems}}
\newtheorem{fact}{Fact}
\newcommand{\bfact}{\begin{fact}}
\newcommand{\efact}{\end{fact}}
\newtheorem{examp}{Example}
\newcommand{\bexamp}{\begin{examp}\rm}
\newcommand{\eexamp}{\end{examp}}
\newtheorem{defn}{Definition}
\newcommand{\bdefn}{\begin{defn}\rm}
\newcommand{\edefn}{\end{defn}}

\newtheorem{alg}{Algorithm}
\newcommand{\balg}{\begin{alg}}
\newcommand{\ealg}{\end{alg}}

\newtheorem{prob}{Problem}
\newcommand{\bprob}{\begin{prob}}
\newcommand{\eprob}{\end{prob}}

\newcommand{\bvtm}{\begin{verbatim}}
\newcommand{\bfig}{\begin{figure}}
\newcommand{\efig}{\end{figure}}
\newcommand{\bcen}{\begin{center}}
\newcommand{\ecen}{\end{center}}

\long\def\comment#1{}

\def \n2{{N_0 \over 2}}

\def \h5{\hspace{0.5in}}

\newcommand{\br}{{\utwi{r}}}

\newcommand{\bu}{{\utwi{u}}}

\newcommand{\bw}{{\utwi{w}}}
\newcommand{\bx}{{\utwi{x}}}
\newcommand{\by}{{\utwi{y}}}

\newcommand{\bA}{{\utwi{A}}}
\newcommand{\bB}{{\utwi{B}}}

\newcommand{\bD}{{\utwi{D}}}

\newcommand{\bG}{{\utwi{G}}}
\newcommand{\bH}{{\utwi{H}}}
\newcommand{\bI}{{\utwi{I}}}
\newcommand{\bJ}{{\utwi{J}}}

\newcommand{\bP}{{\utwi{P}}}

\newcommand{\bX}{{\utwi{X}}}

\usepackage{tikz}
\usetikzlibrary{arrows}
\tikzstyle{block}=[draw opacity=0.7,line width=1.4cm]

\usepackage{epsfig,latexsym,amssymb,amsmath,graphics}
\usepackage{graphicx}

\begin{document}
\IEEEoverridecommandlockouts
%
\title{Wireless MIMO Switching with Network Coding}


\author{Fanggang~Wang and Soung~C.~Liew\\
Institute of Network Coding\\
The Chinese University of Hong Kong \thanks{Corresponding author ***}}

\maketitle

\begin{abstract} \label{abs}
In a generic switching problem, a switching pattern consists of a one-to-one mapping from a set of inputs to a set of outputs (i.e., a permutation). We propose and investigate a wireless switching framework in which a multi-antenna relay is responsible for switching traffic among a set of $N$ stations. We refer to such a relay as a MIMO switch. With beamforming and linear detection, the MIMO switch controls which stations are connected to which other stations.   Each beamforming matrix realizes a permutation pattern among the stations. We refer to the corresponding permutation matrix as a switch matrix. By scheduling a set of different switch matrices, full connectivity among the stations can be established. In this paper, we focus on ``fair switching'' in which equal amounts of traffic are to be delivered for all $N(N-1)$ ordered pairs of stations.  In particular, we investigate how the system throughput can be maximized. In general, for large $N$ the number of possible switch matrices (i.e., permutations) is huge, making the scheduling problem combinatorially challenging. We show that for the cases of $N=4$ and $5$, only a subset of $N-1$ switch matrices need to be considered in the scheduling problem to achieve good throughput. We conjecture that this will be the case for large $N$ as well. This conjecture, if valid, implies that for practical purposes, fair-switching scheduling is not an intractable problem. We also investigate MIMO switching with physical-layer network coding in this paper. We find that it can improve throughput appreciably.
\end{abstract}


%
\IEEEpeerreviewmaketitle

\section{Introduction}
Relaying in wireless networks plays a key role in various communication applications \cite{cov79}. The use of relays can extend coverage as well as improve energy efficiency \cite{den09}. In this paper, we study a set-up in which multiple stations communicate with each other via a multi-antenna relay. With beamforming, the relay controls which stations are connected to which other stations. Each beamforming matrix realizes a permutation among the stations. By scheduling a set of different switch matrices, full connectivity among the stations can be established.


Prior work that investigated the set-up of multiple stations exchanging data via a relay includes \cite{den09, moh09, Cui08}, and \cite{Gao09}. Ref. \cite{den09} studied ``pairwise data exchange'', in which stations form pairs, and two stations in a pair exchange data with each other only. Specifically for pairwise data exchange, if station $i$ transmits to station $j$, then station $j$ transmits to station $i$ as well. In \cite{den09}, MIMO relays with different forward strategies were considered. The bounds on the symmetric capacity were presented. Ref. \cite{moh09} also studied pairwise data exchange, but the relay adopts the decode-and-forward strategy only. The diversity-multiplexing tradeoffs under reciprocal and non-reciprocal channels were analyzed. Both \cite{den09} and \cite{moh09} studied the case in which a station communicates with one other station only. In a general setting, a station could have data for more than one station. In this paper, we focus on a uniform traffic setting in which the amounts of traffic from station $i$ to station $j$ are the same for all $i,j\in \{1, \cdots, N\}$, $i \neq j$. We refer to meeting such a uniform traffic requirement as ``fair switching''. Fair switching is realized by scheduling a set of switch matrices. To the best of our knowledge, the framework of fair switching has not been considered in the existing literature.

Refs. \cite{Cui08} and \cite{Gao09} investigated the case of full data exchange, in which all stations want to broadcast their data to all other stations\footnote{Note that full data exchange is also discussed in \cite{den09}. But they consider a single-antenna relay.}. Data transmissions in \cite{Cui08} and \cite{Gao09} can be summarized as follows: in the first slot, all stations transmit to the relay simultaneously; the first slot is followed by multiple slots for downlink transmissions; in each downlink slot, the relay multiplies the signal received in the first time slot by a different beamforming matrix, such that at the end of all downlink slots, all stations receive the broadcast data from all other stations. By contrast, the framework investigated in this paper is more general in that it can accommodate the pure unicast case, the mixed unicast-multicast case, as well as the pure broadcast case as in \cite{Cui08} and \cite{Gao09}. In particular, a station $i$ can have $M_i$ data streams, and each station $j \neq i$ is a target receiver of one of the $M_i$ streams.

In our framework, the MIMO relay serves as a general switch that switches traffic among the stations. We focus on the use of beamforming at the relay and linear detection to realize different connectivity patterns among the stations.
Each beamforming matrix realizes a permutation connectivity among the stations. By scheduling a set of switch matrices, the MIMO switching system can realize any general transmission pattern (unicast, multicast, broadcast, or a mixture of them) among the stations.

Before delving into technical details, we provide a simple example to illustrate the scenario of interest to us here.  Consider a network with three stations, $1$, $2$, and $3$. The traffic flows among them are shown in Fig.~\ref{3node_demand}: station $1$ wants to transmit ``$a$'' to both stations $2$ and $3$; station $2$ wants to transmit ``$b$'' and ``$c$'' to stations $1$ and $3$, respectively; station $3$ wants to transmit ``$d$'' and ``$e$'' to stations $1$ and $2$, respectively. Pairwise data exchange as in \cite{den09} and \cite{moh09} is not effective in this case because when the number of stations is odd, one station will always be left out when forming pairs.  That is, when the number of stations is odd, the connectivity pattern realized by a switch/permutation matrix does not correspond to pairwise communication. Full data exchange is not appropriate either, since in our example, station $2$ (as well as station $3$) transmits different data to the other two stations. Under our framework, the traffic flows among stations can be met as shown in Fig.~\ref{3node_realization}. In the first slot, station $1$ transmits ``$a$'' to station $3$; station $2$ transmits ``$b$'' to station $1$; station $3$ transmits ``$e$'' to station $2$. In the second slot, station $1$ transmits ``$a$'' to station $2$; station $2$ transmits ``$c$'' to station $3$; station $3$ transmits ``$d$'' to station $1$.

In Section III.C, we will present the details on how to realize the switch matrices.
To limit the scope, this paper focuses on the use of amplify-and-forward relaying and zero forcing (ZF) in establishing the permutations among stations. However, we do generalize the ZF method to one that exploits physical-layer network coding \cite{Ahl00}, \cite{Li03}, \cite{Zhang06physical-layernetwork}, \cite{Katti07embracingwireless} for performance improvement.


The rest of the paper is organized as follows: Section II describes the framework of wireless MIMO switching and introduces the ZF relaying method for establishing permutation among stations. A fair switching scheme is proposed in Section III. In Section IV, we generalize the ZF method to one that exploits network coding. In Section V, we propose two enhanced schemes of MIMO switching. Section VI presents and discusses our simulation results. Section VII concludes this paper.

\section{System Description} \label{sec.SystemDes}

\subsection{System Model}
Consider $N$ stations, $S_1, \cdots, S_N$, each with one antenna. The stations communicate via a relay $R$ with $N$ antennas and there is no direct link between any two stations as shown in Fig.~\ref{diagram}. Each time slot is divided into two subslots. The first subslot is for uplink transmissions from the stations to the relay; the second subslot is for downlink transmissions from the relay to the stations. For simplicity, we assume the two subslots are of equal duration. Each time slot realizes a switching permutation, as described below.

Consider one time slot. Let $\bx = \{x_1, \cdots, x_N\}^T$ be the vector representing the signals transmitted by the stations. We assume that all powers (including noise powers) are normalized with respect to the transmit power of a station. Furthermore, all stations use the same transmit power. Thus, $\mathbb{E}\{x_i^2\}=1,\ \forall\ i$. We also assume that $\mathbb{E}\{x_i\}=0,\ \forall\ i$, and that there is no cooperative coding among the stations so that $\mathbb{E}\{x_i x_j\}=0,\ \forall\ i\neq j$.
Let $\by=\{y_1, \cdots, y_N\}^T$ be the received signals at the relay, and $\bu=\{u_1, \cdots, u_N\}^T$  be the noise vector with i.i.d. noise samples following the complex Gaussian distribution, i.e., $u_n \sim \mathcal{N}_c(0, \sigma_r^2)$. Then
\begin{equation} \label{formula_uplink}
\by=\bH_u \bx +\bu,
\end{equation}
where $\bH_u$  is the uplink channel gain matrix.
The relay multiplies $\by$  by a beamforming matrix $\bG$ before relaying the signals. We impose a power constraint on the signals transmitted by the relay so that
\begin{equation} \label{formula_power}
\mathbb{E}\{\|\bG\by\|^2\}=p.
\end{equation}
Combining (\ref{formula_uplink}) and (\ref{formula_power}), we have
\begin{equation}
\mathbb{E}[\bX^H \bH_u^H \bG^H \bG \bH_u \bX + \bu^H \bG^H \bG \bu ] = p.
\end{equation}
This gives
\begin{equation} \label{formula_tr_power}
\text{Tr}[\bH_u^H \bG^H \bG\bH_u ] + \text{Tr}[\bG^H \bG]\sigma _r^2  = p.
\end{equation}
Let $\bH_d$  be the downlink channel matrix. Then, the received signals at the stations in vector form are
\begin{equation}
\br = \bH_d \bG\by + \bw = \bH_d \bG\bH_u \bx + \bH_d \bG\bu  + \bw,
\end{equation}
where $\bw$ is the noise vector at the receiver, with the i.i.d. noise samples following the complex Gaussian distribution, i.e., $w_n \sim \mathcal{N}_c(0, \sigma^2)$.

\subsection{MIMO Switching}

Suppose that the purposes of $\bG$ are to realize a particular permutation without diagonal elements represented by the permutation matrix $\bP$, and to provide signal amplifications for the signals coming from the stations. That is,
\begin{equation} \label{formula_channel}
\bH_d\bG\bH_u=\bA\bP,
\end{equation}
where $\bA=\diag\{a_1,\cdots,a_N\}$ is an ¡°amplification¡± diagonal matrix. Thus,
\begin{equation} \label{formula_receive}
\br = \left[
\begin{array}{*{20}c}
a_1 x_{i_1} \\
\vdots\\
a_j x_{i_j}\\
\vdots\\
a_N x_{i_N}
\end{array}
\right]+ \bA\bP\bH_u^{-1}\bu+\bw,
\end{equation}
where $S_{i_j}$ is the station transmitting to $S_j$  under the permutation $\bP$  (i.e., in row $j$ of $\bP$, element $i_j$ is one, and all other elements are zero).
Define $\hat \br=\bA^{-1}\br$, i.e., station $S_j$ divides its received signal by $a_j$. We can rewrite (\ref{formula_receive}) as
\begin{equation}
\hat \br = \left[
\begin{array}{*{20}c}
x_{i_1} \\
\vdots\\
x_{i_j}\\
\vdots\\
x_{i_N}
\end{array}
\right]+ \bP\bH_u^{-1}\bu+\bA^{-1}\bw,
\end{equation}
Suppose that we require the received signal-to-noise ratio (SNR) of each station to be the same. Let $h_{u,(i,j)}^{(-1)}$  be element $(i,j)$ in $\bH_u^{-1}$. Then
\begin{equation}\label{9}
\sigma_r^2 \sum\limits_k {|h_{u,(i_j ,k)}^{( - 1)} } |^2  + \frac{{\sigma _{}^2 }}{{|a_j^{} |^2 }} = \sigma _e^2, \quad \forall\ j.
\end{equation}
In other words,
\begin{equation} \label{formula_aj}
|a_j^{} | = \sqrt {\frac{{\sigma _{}^2 }}{{\sigma _e^2  - \sigma _r^2 \sum\limits_k {|h_{u,(i_j ,k)}^{( - 1)} } |^2 }}},\quad \forall\ j,
\end{equation}
where $\sigma_e^2$  is the effective noise power of each station under unit signal power (i.e., the noise-to-signal ratio).

Substituting (\ref{formula_channel}) into (\ref{formula_tr_power}), we have
\begin{equation} \label{formula_expand_power}
 \sum\limits_{i,j} {|h_{d,(i,j)}^{( - 1)} } |^2 |a_j|^2  + {\rm{  }}\sigma _r^2 \sum\limits_{i,k} {|\sum\limits_j {h_{d,(i,j)}^{( - 1)} } a_j h_{u,(i_j ,k)}^{( - 1)} } |^2  = p,
\end{equation}
where $h_{d,(i,j)}^{(-1)}$  is element $(i,j)$ in $\bH_d^{-1}$. If we restrict $a_j\ \forall\ j$ to be real values, then plugging (10) into (11) gives
\begin{equation}  \label{formula_expand_power2}
 \sum\limits_{i,j}  \frac{|{h_{d,(i,j)}^{( - 1)} } |^2{\sigma^2 }}{{\sigma _e^2  - \sigma _r^2 \sum\limits_k {|h_{u,(i_j ,k)}^{( - 1)} } |^2 }}  +
\sigma _r^2 \sum\limits_{i,k} {|\sum\limits_j  \frac{{h_{d,(i,j)}^{( - 1)} }h_{u,(i_j ,k)}^{( - 1)}|\sigma| }{\sqrt {{\sigma _e^2  - \sigma _r^2 \sum\limits_k {|h_{u,(i_j ,k)}^{( - 1)} } |^2 }}  }  } |^2  = p,
\end{equation}

\medskip\noindent{\bf\emph{Problem Definition 1}}: Given $\bH_u, \bH_d, p, \sigma^2, \sigma_r^2$ , and a desired permutation $\bP$, solve for $\bG, \sigma_e^2$.

In the following, we will discuss whether the problem is solvable, and if solvable how to solve the problem. For simplicity, we assume $a_j\ \forall\ j$ are restricted to be non-negative real values. For non-negative real $a_j$, $\sigma_e^2 \in (\max\limits_{i,j,k}\{ \sigma _r^2 \sum\limits_k {|h_{u,(i_j ,k)}^{( - 1)} } |^2\}, +\infty)$ according to (\ref{9}). The left hand side (LHS) of (\ref{formula_expand_power2}) has the following limits:
\[
\lim \limits_{\sigma_e^2 \rightarrow +\infty} \text{LHS} = 0,\qquad \lim\limits_{\sigma_e^2 \rightarrow \max\limits_{i,j,k} \{\sigma _r^2 \sum\limits_k {|h_{u,(i_j ,k)}^{( - 1)} } |^2\}_+} \text{LHS} = +\infty.
\]
Thus, there exists a $\sigma_e^2$ that satisfies (\ref{formula_expand_power2}) for a given power $p$. In the SNR regime of interest, the first item of the LHS is generally dominant because it is the power of desired signals. In this case, the LHS of (\ref{formula_expand_power2}) decreases monotonically as $\sigma_e^2$ increases, and there exists a unique $\sigma_e^2$ for which (\ref{formula_expand_power2}) is satisfied. Given this $\sigma_e^2$, a unique $|a_j|$ can be found from (\ref{formula_aj}). Therefore, in the SNR regime of interest, ``Problem 1'' is always solvable and has a unique solution when $a_j\ \forall\ j$ are non-negative real, and $\bH_u$, $\bH_d$ are invertible.

%

\medskip\noindent{\bf\emph{Numerical Method 1}}: There are $N$ equations in (\ref{formula_aj}) and one equation in (\ref{formula_expand_power}). These equations can be used to solve $a_j$  for $j=1,\cdots,N$  and $\sigma_e^2$. After that (\ref{formula_channel}) can be used to find $\bG$ from $\bA$.

%

\section{Fair Switching} \label{sec.STHardDec}

As has been described in the previous section, in each time slot, the stations transmit to one another according to the switch matrix.
In this section, we study a specific scenario in which each station has an equal amount of traffic to be sent to every other station. The data from station $i$ to station $j$ could be different for different $j$, so this is not restricted to the multicast or broadcast setting. We refer to this setting as "fair switching". To achieve fair switching, multiple transmissions using a succession of different switch matrices will be needed. We next discuss the set of switch matrices we need to construct "fair switching".

\subsection{Derangement} \label{sec.STHardDec.existing}

A derangement is a permutation in which $i$ is not mapped to itself \cite{Has03}. While the number of distinct permutations with $N$ stations is $N!$, the number of derangements is given by a recursive formula
\begin{equation}
d_N = N\cdot d_{N-1}+(-1)^N \triangleq !N ,
\end{equation}
where $d_1=0$, and $!N$ is subfactorial calculated by
\begin{equation}
!N = (N-1)[!(N-1)+ !(N-2)],
\end{equation}
where $!0=1$, $!1=0$. For example, $d_4=9$ although the number of permutations is $4!=24$.
The nine derangements are listed as follows\footnote{Note that for pairwise data exchange \cite{den09, moh09}, pairs of stations want to send data to each other. This corresponds to a symmetric derangement in which element $(i,j) = 1$ implies element $(j,i) = 1$, such as $\bP_1$, $\bP_5$ and $\bP_9$.}:
\[
\begin{array}{l}
\bP_1 = \qquad \qquad \qquad\ \bP_2 = \qquad \qquad \qquad\ \bP_3 = \\
\left[ {\begin{array}{*{20}c}
   0 & 0 & 0 & 1  \\
   0 & 0 & 1 & 0  \\
   0 & 1 & 0 & 0  \\
   1 & 0 & 0 & 0  \\
\end{array}} \right],\quad
\left[ {\begin{array}{*{20}c}
   0 & 0 & 0 & 1  \\
   0 & 0 & 1 & 0  \\
   1 & 0 & 0 & 0  \\
   0 & 1 & 0 & 0  \\
\end{array}} \right],\quad
\left[ {\begin{array}{*{20}c}
   0 & 0 & 0 & 1  \\
   1 & 0 & 0 & 0  \\
   0 & 1 & 0 & 0  \\
   0 & 0 & 1 & 0  \\
\end{array}} \right],
\end{array}
\]
\[
\begin{array}{l}
\bP_4 = \qquad \qquad \qquad\ \bP_5 = \qquad \qquad \qquad\ \bP_6 = \\
\left[ {\begin{array}{*{20}c}
   0 & 0 & 1 & 0  \\
   0 & 0 & 0 & 1  \\
   0 & 1 & 0 & 0  \\
   1 & 0 & 0 & 0  \\
\end{array}} \right],\quad
\left[ {\begin{array}{*{20}c}
   0 & 0 & 1 & 0  \\
   0 & 0 & 0 & 1  \\
   1 & 0 & 0 & 0  \\
   0 & 1 & 0 & 0  \\
\end{array}} \right],\quad
\left[ {\begin{array}{*{20}c}
   0 & 0 & 1 & 0  \\
   1 & 0 & 0 & 0  \\
   0 & 0 & 0 & 1  \\
   0 & 1 & 0 & 0  \\
\end{array}} \right],
\end{array}
\]
\[
\begin{array}{l}
\bP_7 = \qquad \qquad \qquad\ \bP_8 = \qquad \qquad \qquad\ \bP_9 = \\
\left[ {\begin{array}{*{20}c}
   0 & 1 & 0 & 0  \\
   0 & 0 & 1 & 0  \\
   0 & 0 & 0 & 1  \\
   1 & 0 & 0 & 0  \\
\end{array}} \right],\quad
\left[ {\begin{array}{*{20}c}
   0 & 1 & 0 & 0  \\
   0 & 0 & 0 & 1  \\
   1 & 0 & 0 & 0  \\
   0 & 0 & 1 & 0  \\
\end{array}} \right],\quad
\left[ {\begin{array}{*{20}c}
   0 & 1 & 0 & 0  \\
   1 & 0 & 0 & 0  \\
   0 & 0 & 0 & 1  \\
   0 & 0 & 1 & 0  \\
\end{array}} \right].
\end{array}
\]
It can be shown that $\lim_{N\rightarrow \infty}\frac{d_N}{N!}=e^{-1}$ and the limit is approached quite quickly. Thus, the number of derangements is in general very large for large $N$. Performing optimization over this large combinatorial set of derangements in our problem is a formidable task. For example, in our fair switching problem, we want to maximize the system throughput by scheduling over a subset of derangements. It would be nice if for our problem, the optimal solution is not very sensitive to the particular selection of derangements. In Part B, we will formalize the concept of ``condensed derangement sets''.

\subsection{Condensed Derangement Set}

\medskip\noindent{\bf\emph{Definition 1}}: A set of $N-1$ derangements, $\bD_1$, $\bD_2$, $\cdots$, $\bD_{N-1}$, is said to be a \emph{condensed derangement set} if
\begin{equation} \label{formula_condensed_requirement}
\sum\limits_{n=1}^{N-1} \bD_{n}=\bJ-\bI,
\end{equation}
where $\bJ$ is a matrix with all ``1'' elements, and $\bI$ is the identity matrix.


With the help of a computer program, we obtain all the four condensed derangement sets for $N=4$, i.e., $\mathbb{Q}_1=\{\bP_1, \bP_5, \bP_9\}$, $\mathbb{Q}_2=\{\bP_1, \bP_6, \bP_8\}$, $\mathbb{Q}_3=\{\bP_2, \bP_4, \bP_9\}$, and $\mathbb{Q}_4=\{\bP_3, \bP_5, \bP_7\}$.
Furthermore, there are $d_5=44$ derangements for $N=5$ and the number of condensed derangement sets is $56$.

In fair switching, we want to switch an equal amount of traffic from any station $i$ to any station $j$, $i \neq j$. This can be achieved by scheduling the derangements in the condensed derangement set in a weighted round-robin manner (as detailed in ``Approach to Problem 2'' below). Given a condensed set, the scheduling to achieve fair switching is rather simple. However, unless proven otherwise, different condensed sets may potentially yield solutions of different performance. And the number of condensed derangement sets could be huge for large $N$. We define a problem as follows.

\medskip\noindent{\bf\emph{Problem Definition 2}}: Suppose that we want to send equal amounts of traffic from $S_i$  to $S_j$ $\forall\ i\neq j$. Which condensed derangement sets should be used to schedule transmissions? Does it matter?

\medskip\noindent{\bf\emph{Approach to Problem 2}}: The derangements in a condensed derangement set are the building blocks for scheduling. For example, in a complete round transmissions, we may schedule derangement $\bD_n$  for $k_n$  time slots. Then the length of the complete round transmissions will be $\sum\nolimits_{n=1}^{N-1} k_{n}$.

Consider the case of $N=4$. There are four condensed derangement sets. The question is which condensed derangement set will result in the highest throughput. To answer this question, we can approach the problem as follows.

Let $\mathbb{Q}_m = \{\bD_1^m, \bD_2^m, \cdots, \bD_{N-1}^m\}$ be a particular condensed derangement set. For each $\bD_n^m$ , we use ``Numerical Method 1'' above to compute the corresponding $\sigma_e^2$, denoted by $\sigma_{e,n,m}^2$. The Shannon rate is then
\begin{equation}
r_{n,m}=\log (1+\frac{1}{\sigma_{e,n,m}^2}).
\end{equation}
Because of the uniform traffic assumption, we require
\begin{equation}
k_{n,m} r_{n,m} = c,\quad \forall\ n \in [1, \cdots, N-1],
\end{equation}
for some $c$. That is, $c$ is the amount of traffic delivered from one station to another station in one round of transmissions. The effective throughput per station (i.e., the amount of traffic from a station to all other stations) is
\begin{equation} \label{formula_thr}
T_m =  \frac{(N-1)c}{\sum\nolimits_{n=1}^{N-1} k_{n,m}} = \frac{N-1}{\sum_{n=1}^{N-1} 1/r_{n,m}}.
\end{equation}
Numerically, we could first solve for $r_{n,m}$ $\forall\ n$. Then, we apply (\ref{formula_thr}) to find the throughput.

The question we want to answer is whether $T_m$  for different $\mathbb{Q}_m$  are significantly different.
For the case of $N=4$ and $5$, we will show some simulation results indicating that the throughputs of different $\mathbb{Q}_m$ are rather close, and therefore it does not matter much which $\mathbb{Q}_m$ we use.


\subsection{Generalization}

In multi-way relay networks, as mentioned in the introduction, most prior works focus on two patterns of transmissions. The first is pairwise unicast, in which stations form pairs, and the two stations of a pair only communicate with each other \cite{den09, moh09}. The second is the full data exchange, in which each station needs to broadcast to all the other stations \cite{Cui08, Gao09}. In practice, however, the actual transmission patterns could be different from these two patterns. For example, in a video conference session, a subset of stations within the network forms a multicast group, and the transmission pattern is somewhere between the two extremes above.

More generally, in the same network, there could be the co-existence of broadcast sessions, multicast sessions, pairwise unicast sessions, and unidirectional unicast sessions. The MIMO switching framework here is flexible and encompasses this generality.
A scheme for the full data exchange (broadcast) traffic pattern is given in \cite{amah09}. It turns out that we could expand the idea to cater to the general traffic pattern containing a mixture of unicast, multicast, and broadcast as well. For easy explanation, our previous discussion in Part B has an implicit assumption (focus) that each station $i$ wants to send different data to each other station $j \neq i$. If we examine the scheme carefully, this assumption is not necessary. We note that under the scheme, a station will have chances to transmit to all other stations. In particular, a station $i$ will have chances to transmit data to two different stations $j$ and $k$ in two different derangements. If so desired, station $i$ could transmit the same data to stations $j$ and $k$ in the two derangements. This observation in turn implies that the general traffic pattern can be realized.

For illustration, let us examine how the traffic pattern of Fig.~\ref{3node_demand} can be realized. This example is a pattern consisting of the co-existence of unicast and broadcast. As has been described the data transmission can be realized by scheduling a condensed derangement set, which is:
\[
\begin{array}{l}
\bD_1 = \left[ {\begin{array}{*{20}c}
   0 & 1 & 0  \\
   0 & 0 & 1  \\
   1 & 0 & 0  \\
\end{array}} \right],\quad
\bD_2 = \left[ {\begin{array}{*{20}c}
   0 & 0 & 1  \\
   1 & 0 & 0  \\
   0 & 1 & 0  \\
\end{array}} \right].
\end{array}
\]
The transmitted data of station 1, 2 and 3 are respectively $[a, b, e]^T$ for $\bD_1$ and $[a, c, d]^T$ for $\bD_2$.

\section{MIMO Switching with Network Coding} \label{sec.NetCod}

The MIMO switch so far makes use of ZF detection, in which data are switched according to the derangement matrix. In other words, the elements of ``1'' denote the switching desired, while the elements of ``0'' are the interferences that have been forced to zero. A non-zero diagonal element corresponds to self transmission, meaning the relay forwards the self information back to the transmitter. In general, there is no need to force a diagonal element to zero because the self information is known by the transmitter, and can be subtracted out by the transmitter. This is the basic idea behind network coding. Thus, the zero-forcing requirement for the transmission pattern matrix can be relaxed for the diagonal. With this relaxation, there is an extra degree of freedom in our design to improve the system performance.

In this case we rewrite (6) as
\begin{equation} \label{formula_G2}
\bH_d\bG\bH_u=\bA(\bP+\bB),
\end{equation}
where $\bB$ is a diagonal matrix $\bB = \diag\{b_1, \cdots, b_N\}$.

For a symmetric derangement such as $\bP_1$, the corresponding network-coded switch matrix has the following pattern:
\[
\bP^s = \left[ {\begin{array}{*{20}c}
   b_1 & 0 & 0 & 1  \\
   0 & b_2 & 1 & 0  \\
   0 & 1 & b_3 & 0  \\
   1 & 0 & 0 & b_4  \\
\end{array}} \right].
\]
This particular switching pattern corresponds to pairwise data exchange, in which stations 1 and 4 are a pair and stations 2 and 3 are the other pair. Each pairwise transmission is actually a two-way relay channel. Hence, the network coding-based MIMO switching constructs multiple parallel two-way relay transmissions. However, for an asymmetric derangement such as $\bP_2$, the corresponding switch matrix is
\[
\bP^a = \left[ {\begin{array}{*{20}c}
   b_1 & 0 & 0 & 1  \\
   0 & b_2 & 1 & 0  \\
   1 & 0 & b_3 & 0  \\
   0 & 1 & 0 & b_4  \\
\end{array}} \right].
\]
This is not a traditional physical-layer network coding setting \cite{Zhang06physical-layernetwork} because the data exchange is not pairwise. For both symmetric and asymmetric switch matrices, we shall refer to the corresponding matrices with non-zero diagonal as network-coded switch matrix, and the associated MIMO switching setup as MIMO switching with network coding. Note that the concept here generalizes the concept of physical-layer network coding for two-way relay channel in \cite{Zhang06physical-layernetwork}.

Extending the previous ZF method, ZF with network coding yields the following signal after detection.
\begin{equation}
\tilde \br = \left[
\begin{array}{*{20}c}
x_{i_1} \\
\vdots\\
x_{i_j}\\
\vdots\\
x_{i_N}
\end{array}
\right] + \bA^{-1}\bB\bx + (\bP+\bB)\bH_u^{-1}\bu+\bA^{-1}\bw.
\end{equation}
The self-information term $\bA^{-1}\bB\bx$ can be cancelled at the stations. Potentially, with network coding, we could decrease the noise power by designing $\bB$ appropriately. The effective noise power can be written by
\begin{equation}
\sigma_r^2 \sum\limits_k {|h_{u,(i_j ,k)}^{( - 1)} } |^2 +  \sigma_r^2 |b_j|^2 \sum\limits_k {|h_{u,(j ,k)}^{( - 1)} } |^2 + \frac{{\sigma _{}^2 }}{{|a_j^{} |^2 }} = \sigma _e^2, \quad \forall j,
\end{equation}
that is
\begin{equation} \label{formula_aj2}
|a_j^{} | = \frac{{|\sigma| }}{\sqrt {{\sigma _e^2  - \sigma _r^2 \sum\limits_k {|h_{u,(i_j ,k)}^{( - 1)} } |^2 - \sigma_r^2 |b_j|^2 \sum\limits_k {|h_{u,(j ,k)}^{( - 1)} } |^2 }}},\quad \forall j.
\end{equation}
The relay transmission power can be rewritten as
\begin{equation} \label{formula_power2}
 \text{Tr}\{[\bH_d^{ - 1} \bA(\bP+\bB)]^H \bH_d^{ - 1} \bA(\bP+\bB)\} +
 \sigma _r^2 \cdot \text{Tr}\{[\bH_d^{ - 1} \bA(\bP+\bB)\bH_u^{ - 1} ]^H \bH_d^{ - 1} \bA(\bP+\bB)\bH_u^{ - 1} \}  = p. \end{equation}

\medskip\noindent{\bf\emph{Problem Definition 3}}: Given $\bH_u, \bH_d, p, \sigma^2, \sigma_r^2$ , and a desired permutation $\bP$. Solve for $\bG,\ \bB,\ \sigma_e^2$ to optimize the effective throughput of the network.

\medskip\noindent{\bf\emph{Numerical Method 3}}: We give a naive method here. Assume $\bB = b\bI$ and $b$ is a real scalar for simplicity. We set a search range for $b$, and for each trial of $b$, there are $N$ equations in (\ref{formula_aj2}) and one equation in (\ref{formula_power2}). Use the $N+1$ equations to solve $a_j$  for $j=1,\cdots,N$  and $\sigma_e^2$. Then find the minimum $\sigma_e^2$ in among all trials within the search range of $b$, and the corresponding $\bA$ and $\bB$ as the solution.

\section{Discussion on Complex $\bA$ and $\bB$}
$\bA$ and $\bB$ are restricted to be real so far. In this section we consider complex $\bA$ and $\bB$. Recall that with a given effective noise power $\sigma_e^2$ we can calculate the power consumption of the relay. For complex $\bA$ and $\bB$, with the extra degree of freedom of their phases, the relay power consumption can be reduced. In other words, given the transmit power constraint of the relay we can optimize the phases of $\bA$ and $\bB$ to reduce $\sigma_e^2$. However, the problem is nonconvex thus intractable.

We propose two schemes to exploit the phases of $\bA$ and $\bB$. The first scheme is targeted for the pairwise transmission pattern without network coding. The amplitudes of $\bA$ are calculated by Numerical Method 1. Then for any pair of stations in the overall pairwise pattern, their elements in $\bA$ are set to have counter phases (i.e., out of phase by $\pi$). This setting could be proved to be optimal for two-way relay channel. To limit the scope of the current paper, this new result of ours will be published shortly elsewhere. For larger numbers of stations, this setting can still improve the throughput performance, as will be shown in Section VI. The second scheme is for the general case (pairwise or non-pairwise transmission pattern, with or without network coding). The amplitudes of $\bA$ and $\bB$ are calculated by Numerical Method 1 and 3 respectively. Then we divide the interval of $[0,2\pi)$ equally into $M$ bins with the values of $0,\frac{2\pi}{M},\cdots,\frac{2(M-1)\pi}{M}$ respectively and randomly pick among them to set the phase values for the variables of $\bA$ and $\bB$. We perform $L$ trials of these random phase assignments. For each trial we could solve for $\sigma_e^2$ by Numerical Method 3, and then compare the $L$ solutions of $\sigma_e^2$ and select the trial which has the minimum $\sigma_e^2$. Thus, the phase assignment of this trial is used to set the phase values of $\bA$ and $\bB$. In Section VI, we will show that large gains can be achieved with only a small number of bins and trials.


%

\section{Simulation}
In this section, we evaluate the throughputs of different schemes. We assume that the uplink channel $\bH_u$ and downlink channel $\bH_d$ are reciprocal, i.e., $\bH_d=\bH_u^T$, and they both follow the complex Gaussian distribution $\mathcal{N}_c(\textbf{0},\bI)$. Thus, elements of the channels are i.i.d. and follow the complex Gaussian distribution $\mathcal{N}_c(0,1)$. We assume the relay has the same transmit power as all the stations, i.e., $p=1$.

First, we answer the question raised in Problem Definition 2. We analyze the scenarios where $N=4$ and $N=5$, and use the random-phase scheme described in Section V to calculate $\bA$. The four different condensed derangement sets of $N=4$, $\mathbb{Q}_1$, $\mathbb{Q}_2$, $\mathbb{Q}_3$ and $\mathbb{Q}_4$, are considered for fair switching.
For each channel realization, we evaluate the throughput per station, $T_m$, as given by (\ref{formula_thr}). We simulated a total of 10000 channel realizations and computed the expected throughput averaged over the channel realizations. We find that the four condensed derangement sets yield essentially the same average throughput (within 1\% in the medium and high SNR regimes and within 3\% in the low SNR regime). Fig.~\ref{setsele} plots the throughput for one of the condensed derangement set. For $N=5$ there are $56$ different condensed derangement sets. As with the $N=4$ case, all the sets have roughly the same average throughput (within 1\%). Fig.~\ref{setsele} plots the result of one set.
We conjecture that different condensed derangement sets achieve roughly the same average throughput for $N$ larger than $5$ as well. A concrete proof remains an open problem. The ramification of this result, if valid, is as follows. For large $N$, the number of condensed derangement set is huge, and choosing the optimal set is a complex combinatorial problem. However, if their relative performances do not differ much, choosing any one of them in our engineering design will do, significantly simplifying the problem.

We next evaluate MIMO switching with and without network coding in single slot, i.e., for one derangement. First, we consider real variables. A scheme is proposed in the literature \cite{amah09}, which investigates a similar problem as ours. It simply uses a positive scalar weight to control the relay power consumption instead of our diagonal $\bA$. As shown in Fig.~\ref{thr_cmp}, it has almost the same average throughput as our MIMO switching scheme with real $\bA$ and without network coding. Henceforth we regard the MIMO switching scheme with real $\bA$ and without network coding as a benchmark and call it ``the basic scheme''. Despite the same average throughput performance, the basic scheme has an advantage over the scalar scheme in \cite{amah09} in that the basic scheme guarantees fairness. That is, in our basic scheme, each station has exactly the same throughput, while the stations in the scheme in \cite{amah09} could have varying throughputs. The scalar scheme in \cite{amah09} focuses on optimizing the sum rate of all stations; the individual rates of the stations may vary widely with only one degree of freedom given by the scalar. Fig.~\ref{thr_cmp} also shows that when both $\bA$ and $\bB$ are real, our MIMO switching with network coding has 0.5dB gain compared to the basic scheme.

Next, we consider complex $\bA$ without network coding. The random-phase scheme proposed in Section V with 10 and 100 trials are shown in Fig.~\ref{thr_cmp}, and the two curves achieve 0.6dB and 0.7dB gains respectively over the basic scheme. Thus, the scheme with complex $\bA$ can potentially achieve even larger gain than the network coding scheme with real $\bA$ and $\bB$. The other proposed scheme in Section V is for pairwise transmission pattern. We set the two corresponding $\bA$'s elements in one pair to have counter phases. Its average throughput is in between those of the random-phase schemes with 10 and 100 trials, and achieves 0.65dB gain over the basic scheme. The advantage of the  counter-phase scheme is that it has low complexity comparable to the basic scheme.

However, for non-pairwise transmission, as shown in Fig.~\ref{thr_cmp2} the counter-phase scheme cannot achieve any gain over the basic scheme, while random-phase scheme still can.

Lastly, we consider complex $\bA$ and $\bB$ for MIMO switching with network coding. We evaluate the random-phase scheme with 10 and 100 trials, and they can achieve 1.5dB and 2.2dB gain over the basic scheme. Therefore, with network coding the throughput performance can be further improved compared with the other MIMO switching schemes.

To sum up this section, three general results are stated as follows:

\medskip\noindent{\bf\emph{General Result 1}}: In the framework of MIMO fair switching with $4$ or $5$ stations, any condensed derangement set can be used to schedule a fair switching because different condensed derangement sets achieve roughly the same average throughput. We conjecture that this will be the case when $N$, the number of stations, is large as well. If this conjecture holds, then the issue of condensed set selection will go away, and the complexity of the optimization problem will be greatly reduced. This is especially so when $N$ is large because the number of different condensed derangement sets grow exponentially with $N$.

\medskip\noindent{\bf\emph{General Result 2}}: Network coding can be applied in MIMO switching to significantly improve average throughput performance. It is worth mentioning that network coding helps not only for the traditional pairwise switching pattern but also for the non-pairwise pattern. The non-pairwise pattern has not been treated in the existing literature prior to this paper.

\medskip\noindent{\bf\emph{General Result 3}}: For pairwise switch without network coding, the counter-phase scheme outperforms the basic scheme with real $\bA$. Furthermore, the counter-phase scheme has low complexity because its computation does not optimize over the phases of $\bA$.

\section{Conclusions}\label{sec.Conclusion}

We have proposed a framework for wireless MIMO switching to facilitate communications among multiple wireless stations. In our framework, a multi-antenna relay controls which stations are connected to which other stations with beamforming. Specifically each beamforming matrix at the multi-antenna relay realizes a switching permutation among the stations, represented by a switch matrix. By scheduling a set of switch matrices, full connectivity among the stations can be established. A salient feature of our MIMO switching framework is that it can cater to general traffic patterns consisting of a mixture of unicast traffic, multicast traffic, and broadcast traffic flows among the stations.

There are many nuances and implementation variations arising out of our MIMO switching framework. In this paper, we have only evaluated the performance of some of the variations. We first studied the ``fair switching'' setting in which each station wants to send equal amounts of traffic to all other stations. In this setting, we aim to deliver the same amount of data from each station $i$ to each station $j \neq i$ by scheduling a set of switch matrices. In general, many sets of switch matrices could be used for such scheduling. The problem of finding the set that yields optimal throughput is a very challenging problem combinatorially. Fortunately, for number of stations $N = 4$ or $5$, our simulation results indicate that different sets of switch matrices achieve roughly the same throughput, essentially rendering the selection of the optimal set a non-issue. We conjecture this will be the case for larger $N$ as well. If this conjecture holds, then the complexity of the optimization problem can be decreased significantly as far as engineering design is concerned.

We next moved to the study of single switch matrix and investigated the performances of different realizations for the same switch matrix (i.e., realizations using different beamforming matrices). Our general conclusion is that the use of physical-layer network coding can improve the throughput performance appreciably. In addition, we discover an interesting result for pairwise switch matrices. The computation cost of the beamforming matrix could be high in general. However, when the switch matrix is pairwise and network coding is not used, the computation of the beamforming matrix can be much simplified with a ``counter-phase'' approach.

There are many future directions going forward. For example, the beamforming matrices used in our simulation studies could be further optimized. In addition, the setting in which there are unequal amounts of traffic between stations will be interesting to explore. Also, this paper has only considered switch matrices those realize full permutations in which all stations participate in transmission and reception in each slot; it will be interesting to explore switch matrices that realize connectivities among stations that are not a full permutation. Finally, future work could also explore the case where the number of antenna at the relay is not exactly $N$.

\bibliographystyle{IEEEtran}
\bibliography{MIMO_switch}

\newpage

\tikzstyle{place}=[circle,draw=blue!50,fill=blue!20,thick,
inner sep=0pt,minimum size=6mm]
\tikzstyle{transition}=[rectangle,draw=black!100,fill=none,semithick,
inner sep=0pt,minimum size=6mm]
\tikzstyle{mimoswitch}=[rectangle,draw=black!100,fill=none,semithick,
inner sep=0pt,minimum width=12mm, minimum height=26mm]
\tikzstyle{relay}=[rectangle,draw=black!100,fill=none,semithick,
inner sep=0pt,minimum width=25mm, minimum height=10mm]

\def\antenna{%
    -- +(0mm,4.0mm) -- +(2.625mm,7.5mm) -- +(-2.625mm,7.5mm) -- +(0mm,4.0mm)
}

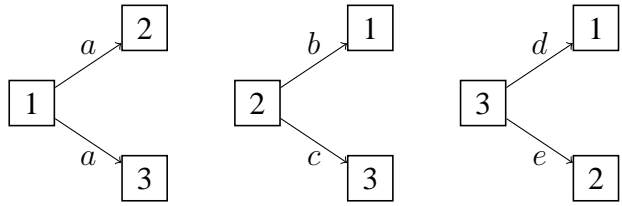
\begin{figure}
\centering
\begin{tikzpicture}
\node (1s1) at (-1,1) [transition] {1};
\node (1s2) at ( .5,2) [transition] {2};
\node (1s3) at ( .5,0) [transition] {3};
\path[->] (1s1) edge node  [above]      {$a$} (1s2)
                edge node  [below] {$a$} (1s3);
\node (2s2) at ( 2,1) [transition] {2};
\node (2s1) at ( 3.5,2) [transition] {1};
\node (2s3) at ( 3.5,0) [transition] {3};
\path[->] (2s2) edge node [above]{$b$} (2s1)
                edge node [below]{$c$} (2s3);
\node (3s3) at ( 5,1) [transition] {3};
\node (3s1) at ( 6.5,2) [transition] {1};
\node (3s2) at ( 6.5,0) [transition] {2};
\path[->] (3s3) edge node [above] {$d$} (3s1)
                edge node [below] {$e$} (3s2);
\end{tikzpicture}
\caption{Traffic demand of a three stations example.}
\label{3node_demand}
\end{figure}

\begin{figure}
\centering
\begin{tikzpicture}[x=2.2em,y=2.8em]
\node (1s1) at (-1,2) [transition] {1};
\node (1s2) at (-1,1) [transition] {2};
\node (1s3) at (-1,0) [transition] {3};

\node (2s3) at (3,2) [transition] {3};
\node (2s1) at (3,1) [transition] {1};
\node (2s2) at (3,0) [transition] {2};

\node (m1) at (1,1) [mimoswitch] {\parbox{12mm}{MIMO \\ switch}};
\node (m2) at (7,1) [mimoswitch] {\parbox{12mm}{MIMO \\ switch}};

\node (3s1) at (5,2) [transition] {1};
\node (3s2) at (5,1) [transition] {2};
\node (3s3) at (5,0) [transition] {3};

\node (4s2) at (9,2) [transition] {2};
\node (4s3) at (9,1) [transition] {3};
\node (4s1) at (9,0) [transition] {1};

\path[->] (1s1) edge node [above]{$a$} (m1.west|-1s1);
\path[->] (m1.east|-2s3) edge node [above]{$a$} (2s3);
\path[->] (1s2) edge node [above]{$b$} (m1.west|-1s2);
\path[->] (m1.east|-2s1) edge node [above]{$b$} (2s1);
\path[->] (1s3) edge node [above]{$e$} (m1.west|-1s3);
\path[->] (m1.east|-2s2) edge node [above]{$e$} (2s2);

\path[->] (3s1) edge node [above]{$a$} (m2.west|-3s1);
\path[->] (m2.east|-4s2) edge node [above]{$a$} (4s2);
\path[->] (3s2) edge node [above]{$c$} (m2.west|-3s2);
\path[->] (m2.east|-4s3) edge node [above]{$c$} (4s3);
\path[->] (3s3) edge node [above]{$d$} (m2.west|-3s3);
\path[->] (m2.east|-4s1) edge node [above]{$d$} (4s1);

\node at (1,-1) {Slot 1};
\node at (7,-1) {Slot 2};
\end{tikzpicture}
\caption{A transmission established by two slots of unicast connectivity realizes the traffic demand in Fig.~\ref{3node_demand}.}
\label{3node_realization}
\end{figure}
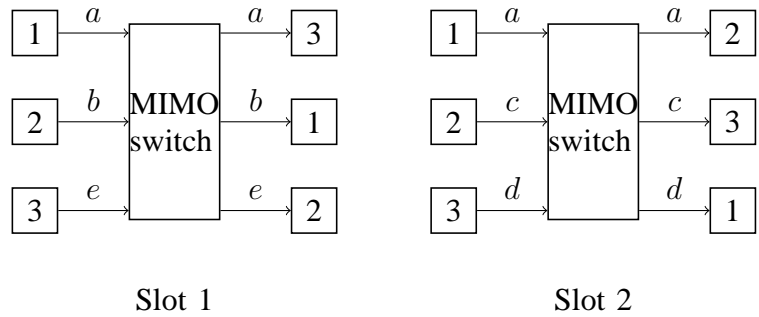

\begin{figure}
\centering
\begin{tikzpicture}[node distance=10em]
\node (relay) [relay] {Relay};
\node (1) [above right of=relay] [transition] {1};
\node (2) [above left of=relay] [transition] {2};
\node (3) [below left of=relay] [transition] {3};
\node (N) [below right of=relay] [transition] {$N$};

\draw[color=black,semithick] (1.north) \antenna;
\draw[color=black,semithick] (2.north) \antenna;
\draw[color=black,semithick] (3.north) \antenna;
\draw[color=black,semithick] (N.north) \antenna;
\path (relay.north west) to node (a) [pos=.2,inner sep=0] {} (relay.north) to node (b) [pos=.8,inner sep=0] {} (relay.north east);
\draw[color=black,semithick] (a) \antenna  (b) \antenna;

\draw [dashed] (-.8,.8) to node [above]  {$N$} (.8,.8);
\draw [dashed] (-1.5,-2.5) to node [auto] {} (1.5,-2.5);
\draw [color=blue, dashed, semithick, ->] (1.3,1.5) -- (1.9,2.1); \draw [semithick, ->] (2.1,1.9) -- (1.5,1.3);
\draw [color=blue, dashed, semithick, ->] (-1.3,1.5) -- (-1.9,2.1); \draw [semithick, ->] (-2.1,1.9) -- (-1.5,1.3);
\draw [color=blue, dashed, semithick, ->] (1.3,-0.8) -- (1.9,-1.4); \draw [semithick, ->] (2.1,-1.2) -- (1.5,-0.6);
\draw [color=blue, dashed, semithick, ->] (-1.3,-0.8) -- (-1.9,-1.4); \draw [semithick, ->] (-2.1,-1.2) -- (-1.5,-0.6);

\draw [color=blue, dashed, semithick, ->] (-2.8,-3.7) -- (-1.7,-3.7); \draw [semithick, ->] (.4,-3.7) -- (1.5,-3.7);
\node at (-0.8,-3.7) {Subslot 1};
\node at (2.4,-3.7) {Subslot 2};
\end{tikzpicture}
\caption{Wireless MIMO switching.}
\label{diagram}
\end{figure}
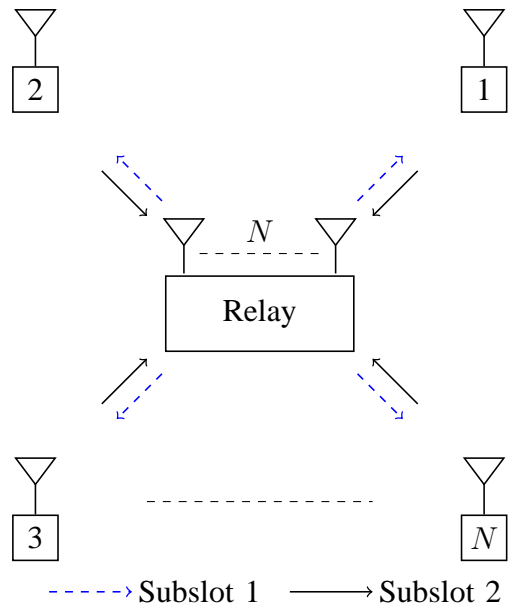


\begin{figure}
\centering
\includegraphics[width=7in]{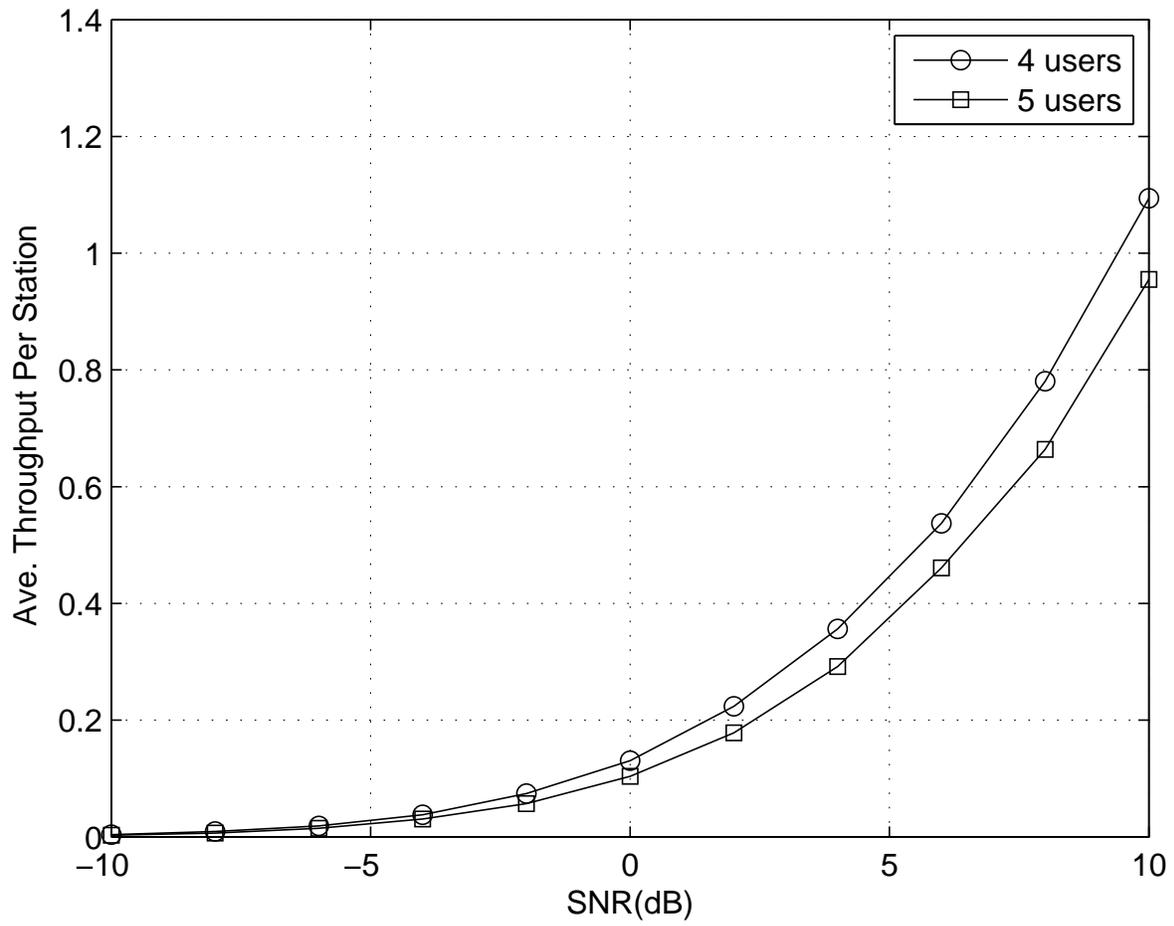}
\caption{Average throughput per station under MIMO fair switching. Only the results of one condensed derangement set is presented because the results of other derangement sets are within 3\% of the results shown here.}
\label{setsele}
\end{figure}

\begin{figure}
\centering
\includegraphics[width=7in]{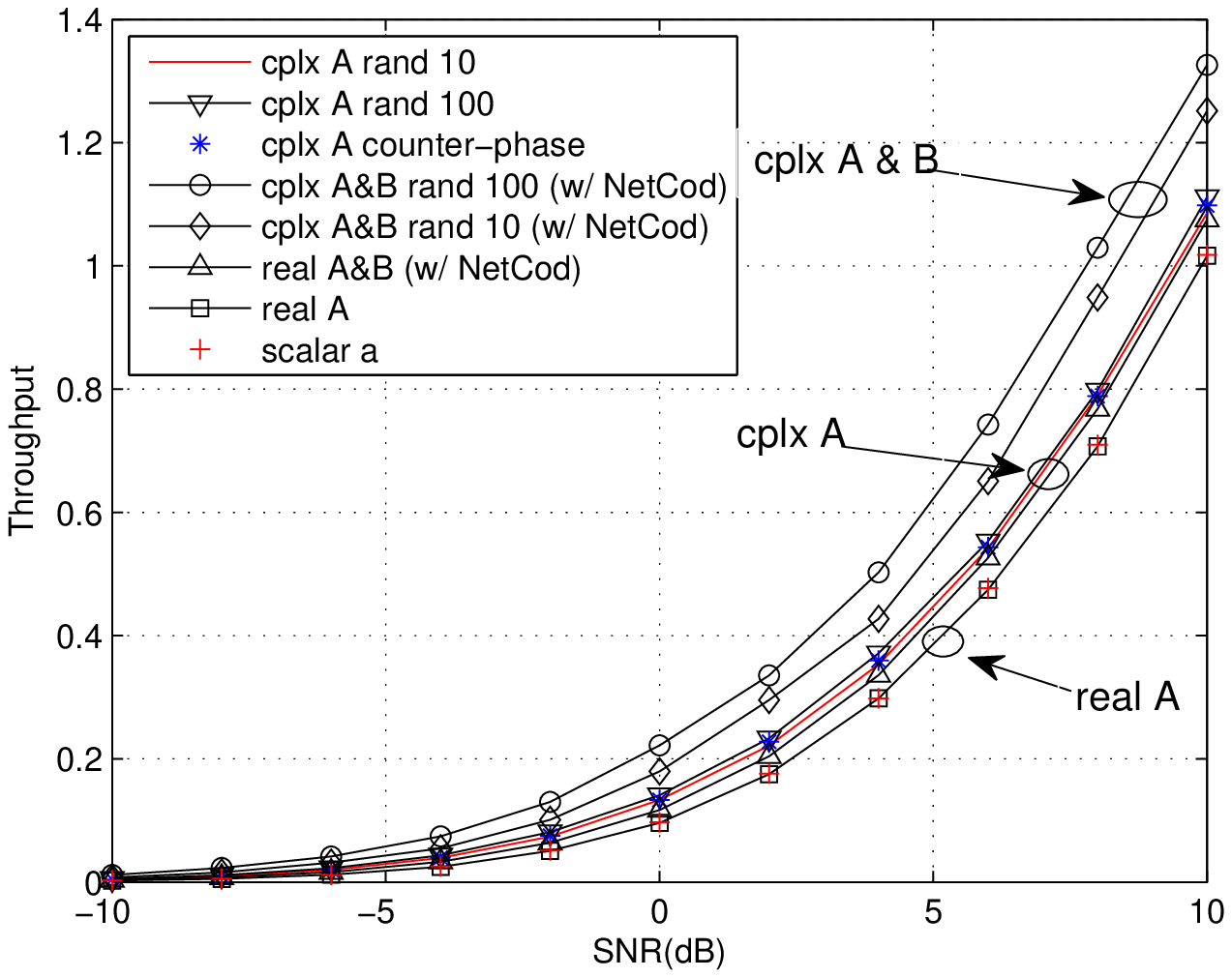}
\caption{Throughput comparison of different MIMO switching schemes for pairwise switching pattern.}
\label{thr_cmp}
\end{figure}

\begin{figure}
\centering
\includegraphics[width=7in]{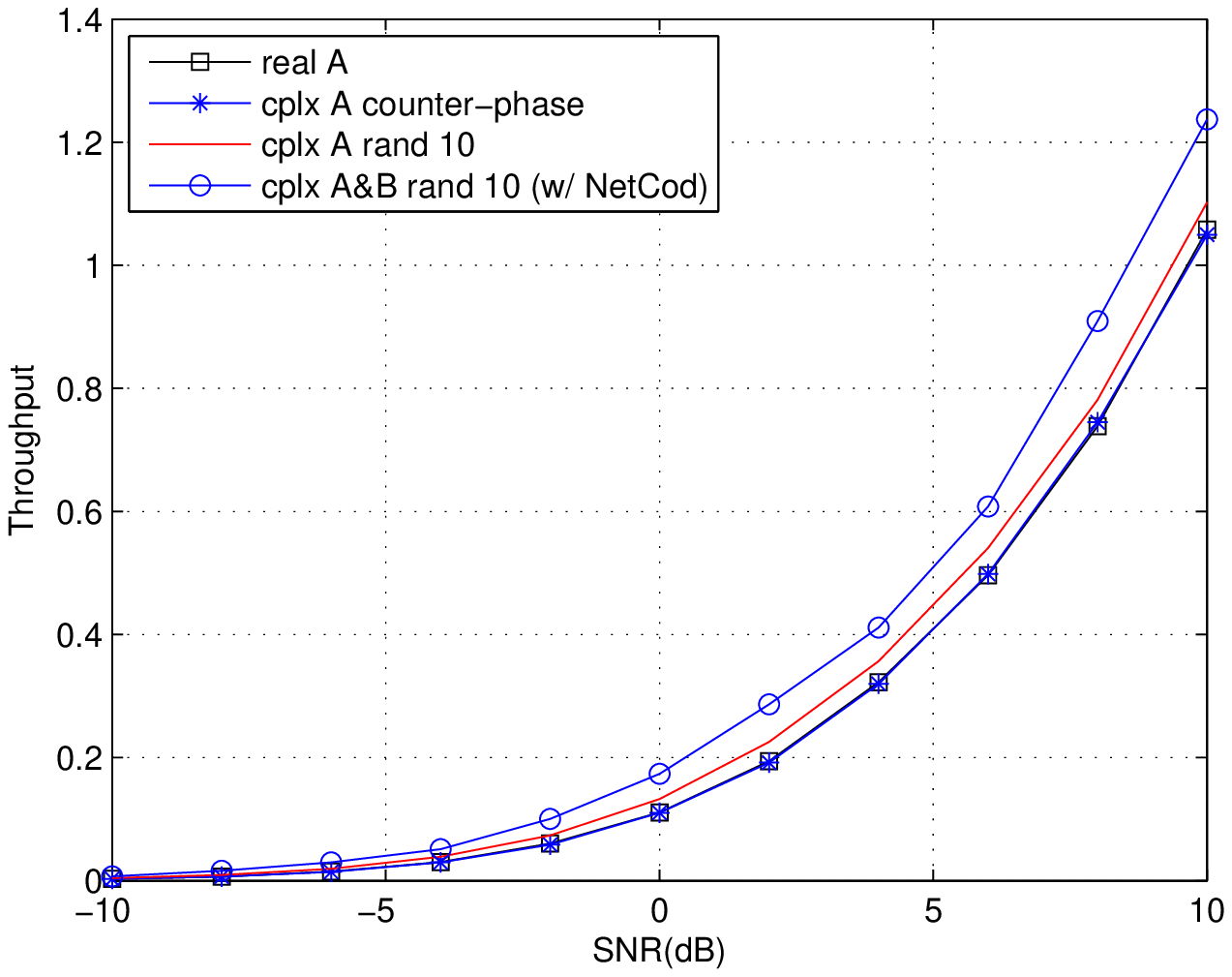}
\caption{Throughput comparison of different MIMO switching schemes for non-pairwise switching pattern.}
\label{thr_cmp2}
\end{figure}

\end{document}